Atomic and Molecular Nitrogen Ions at the Dayside Magnetopause During the 2024 Mother's Day Storm

Gomez, R.G., Fuselier, S. A., Vines, S. K., Goldstein, J., Burch, J. L., Strangeway, R. J.

# 1   Abstract

Ion measurements made with the Hot Plasma Composition Analyzers of the Magnetospheric Multiscale Mission (MMS-HPCAs) during the Mother's Day Storm (Gannon Storm) of 10–13 May 2024 yield the first observations of atomic and molecular nitrogen ions in the Earth's dayside outer magnetosphere. A population of ions identified as doubly charged nitrogen and oxygen was also measured. These observations were made within a highly compressed magnetosphere at a geocentric distance of ~6 $R_E$ during the early recovery phase of the storm. From the ion composition measurements and accompanying magnetic field data, we determine the reconnection rate at the magnetopause; we compare this result to a model reconnection rate that assumes the presence of only atomic oxygen and hydrogen. The heavy-ion-laden mass density in the magnetosphere was greater than the shocked solar wind mass density in the magnetosheath. Despite these conditions, magnetic reconnection still occurred at the magnetopause.

# 2   Key points

- Atomic and molecular nitrogen ions from the ionosphere observed in the dayside magnetopause at ~6 Earth Radii
- The ionospheric ions mass-loaded the plasma and reduced the reconnection rate.
- Acquisitions in the magnetosphere and Low Latitude Boundary Layer allowed $N^+/O^+$ in both.

## 3  Plain Language Summary

Observations during the 2024 Mother's Day Storm revealed the presence of atomic and molecular nitrogen ions in the Earth's dayside outer magnetosphere. This measurement marks the first observation of these ions in the region.

## 4  Introduction and Initial Observations

The energization and transport of nitrogen ions throughout the ionosphere-magnetosphere system is an ongoing research subject. Measurements of upflowing nitrogen ions start as early as 1961 by Sputnik III and span over six solar cycles and include a varied range of scale and locations (cf. review by Ilie et al., 2023, and the references therein).

The Mother's Day Storm (also known as the "Gannon Storm") that occurred from May 10-13, 2024, was driven by a series of coronal mass ejections (CMEs) that directly impacted the Earth's magnetosphere. The Mother's Day Storm, characterized by a minimum DST index of -412 nT, was the most intense geomagnetic storm since the 2003 "Halloween" Storm (Schebbetten et al., 2024). Its most apparent effects were auroral observations well into the lower forty-eight states, including the northern regions of Texas. However, effects not readily visible and only observable with dedicated plasma instruments occurred as well.

Early in the recovery phase of the storm, on May 11, 2024, the Magnetospheric Multiscale Mission observations, made with the Hot Plasma Composition Analyzers (MMS-HPCAs, Young, et al., 2016) measured an unexpectedly high flux of atomic oxygen ions ($O^+$) in the pre-noon magnetosphere at a highly compressed geocentric distance of ~6 $R_E$. Figure 1 provides an overview of the magnetopause crossings during 2024-05-11 14:36 – 14:51 UT and the spacecraft location. Because of the close separation of the four MMS spacecraft (average of ~20 km), the "survey-mode" observations in Figure 1 are nearly identical between the spacecraft, so only MMS1 is shown.

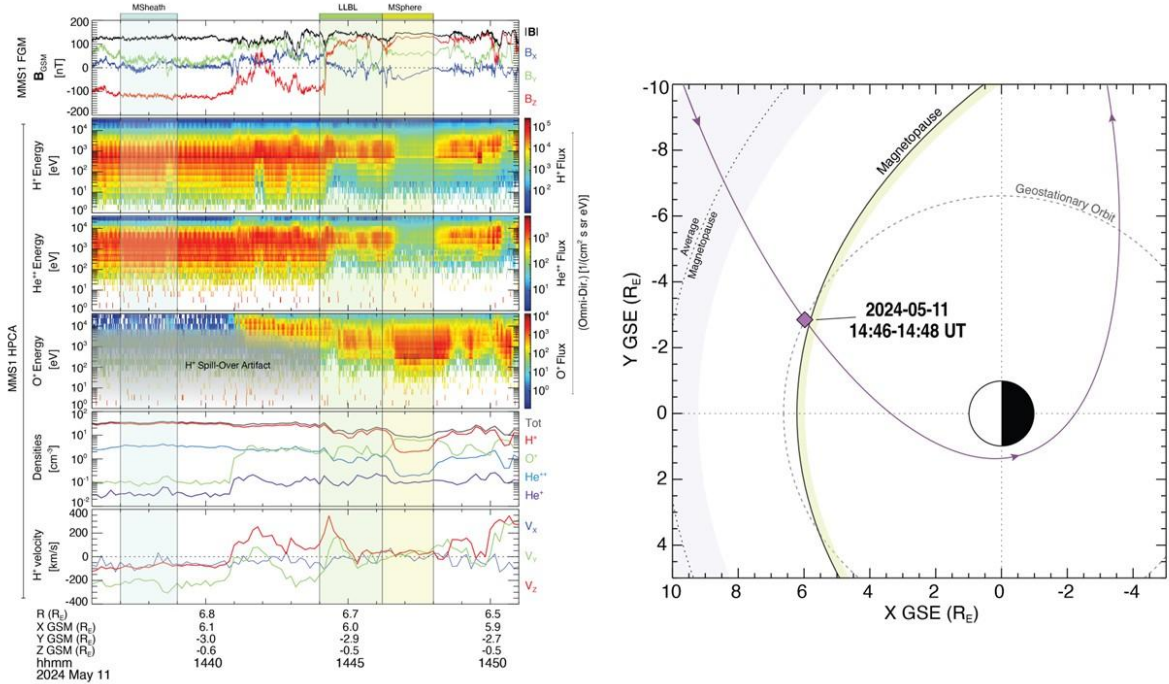

*Figure 1: **Magnetopause crossing 2024-05-11 14:36 - 14:51 UTC.** LEFT: MMS traversed magnetosheath ("MSheath"), the low-latitude boundary layer (LLBL), and magnetosphere ("Msphere"). Data panels: MMS1 FGM magnetic field (GSM coordinates), HPCA $H^+$ and $O^+$ omni-directional flux, HPCA densities (total, $H^+$, $O^+$, $He^{2+}$, and $He^+$ and total (tot)). The magnetosphere encounter (14:46:10 – 14:48 UT) features elevated $O^+$ flux compared to surrounding regions. RIGHT: Orbit of MMS during magnetopause encounter, at geocentric radius ~6 $R_E$. Solid black line: Shue et al. (1998) model magnetopause (OMNI $B_Z$ = **xxx nT**, $P_{dyn}$ = **xx nPa** at 14:46 UT); yellow shaded region is model variance for OMNI data between 14:30-1500 UT. Dashed line, light grey shading: 4-year average and variance.*

As seen in Figure 1, ion measurements, particularly the number densities, in the magnetosphere display a surprisingly high abundance of oxygen ions relative to protons, greater by almost a factor of four. This unexpected difference in the relative ion densities led to further analysis using the HPCA time-of-flight (TOF) data product.

5   TOF Analysis and Determining the Relative Abundances of $N^+$, $O^+$, and $N_2^+$

In this section, more detailed time-of-flight (TOF) analysis reveals the presence of atomic and molecular nitrogen ions ($N^+$ and $N_2^+$) with the enhanced $O^+$. HPCA provides fluxes and velocity distribution functions (VDFs) of five ion species from its primary TOF product: protons, singly and doubly charged helium, atomic oxygen ions, and a "background" at mass/charge = 8 (in proton mass units). The ion VDFs are used for determining plasma moments, such as the ion number densities and velocities shown in Figure 1. After processing, the Burst Resolution TOF product is stored as an array with dimensions 32 E/Q channels by 256 TOF channels (a 1 ns TOF resolution).

Figure 2 shows the collapsed Flux-TOF spectrum from the magnetosphere encounter and the results of a cut at 3.2 keV/e.

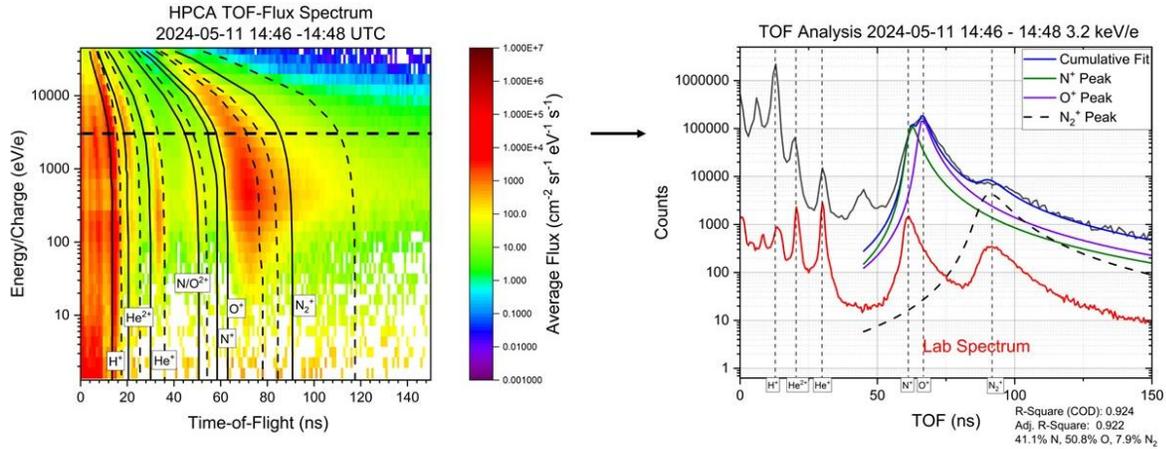

*Figure 2: HPCA TOF-Flux Spectrum 2024-05-11 14:46 - 14:48 UTC & TOF spectrum cut at 3.2 keV/e. LEFT: integrated TOF-Flux Spectrum over the period. Alternating solid & dashed lines indicate the beginning and end TOF ranges for indicated ion species. RIGHT: spectrum cut at 3.2 keV/e. Vertical lines indicate the ion peak location. TOF peak fits for atomic nitrogen, atomic oxygen, and molecular nitrogen are included. Lab Spectrum (red) acquired during HPCA calibration at beam energy 3.159 keV/e.*

In the plot at left, solid and dashed line pairs indicate the start and end of a particular ion's TOF range, respectively. These TOF ranges were determined during instrument calibration (Young et al., 2016). Clear separation is achieved for M/Qs of 1, 2, and 4, but separation becomes more difficult for higher masses, evidenced by the overlapping ranges of atomic nitrogen and oxygen ions (M/Qs of 14 and 16). Overlapping TOF ranges occur for higher mass ions because of energy straggling in the carbon foils. Thus, determining the presence of specific ions requires fitting the TOF spectra at different energies. The magnetospheric Flux-TOF spectrum in Figure 2 displays several interesting features. The most obvious is the significant contribution in the molecular nitrogen ion TOF range ($N_2^+$). Another is the enhanced contributions within the atomic nitrogen & oxygen TOF ranges.

HPCA was calibrated with 0.5 ns TOF resolution and multi-species ion beams. One calibration energy, 3.159 keV/e, is close to the 3.2 keV/e of the cut indicated by the dashed horizontal line at left in Figure 2. The red TOF spectrum in the right plot shows the calibrated HPCA response to a multi-species 3.159 keV/e ion beam produced in the Energetic Plasma Instrument Calibration (EPIC) system (Young et al., 2016). The vertical dashed lines in the plot correspond to the peak centroids of $H^+$, $H_2^+$ (a proxy for $He^{2+}$), $He^+$, $O^{2+}$, $N^+$, and $N_2^+$.

The in-situ TOF spectrum in the same plot shows contributions from expected magnetospheric species ($H^+$, $He^{2+/+}$) and evidence of $O^+$. However, the $O^+$ peak, as seen in the cut, has the correct centroid but is too broad to be oxygen alone. In addition, the oxygen peak tail includes a "shelf" structure explainable only by contributions from more massive ions, likely molecular nitrogen. Also present in the in-situ spectrum is a peak expected for doubly-ionized oxygen, $O^{2+}$ but also too broad to be due to oxygen alone(no doubly-ionized species were produced in the multi-species beam used during lab calibration).

After calibration, the ion-specific TOF ranges shown in Figure 2 (left plot) were determined by fitting single- and multi-species TOF peaks with a Pearson IV function (Pearson, 1895). The Pearson IV function accurately characterizes the leading edge (faster TOF) and extended tails (slower TOF) of the peaks produced in HPCA's carbon foil TOF section. The pre-determined peaks for $N^+$, $O^+$, and $N_2^+$ were used to fit the in-situ spectrum at TOF values greater than 45 ns. The resulting individual and cumulative fits for TOF > 45 are shown in the plot at right in Figure 2.

The combined contributions of the $N^+$, $O^+$, and $N_2^+$ peaks correspond with the cumulative counts in the TOF region of interest (45 ns - 200 ns, $R^2$ =0.924 and adj. $R^2$ = 0.922) and visually adhere to their cumulative appearance. Post-fitting, the relative peak areas were used to determine the relative concentrations of the ions present: 41.1% $N^+$, 50.8% $O^+$, and 7.9% $N_2^+$, with an $N^+/O^+$ of ~0.81.

Atomic and molecular nitrogen are well-known ionospheric species and are not expected to be found within the magnetosphere and at the geocentric distance in this investigation. While their insertion mechanism is unknown, verifying these TOF results marks this instance as the first observation of atomic and molecular nitrogen in the dayside magnetosphere near the equatorial region.

6  Comparing Compositions in the Magnetosphere and the Lower Latitude Boundary Layer

The presence of the nitrogen (atomic and molecular) ions is attributed to geomagnetic storm-generated injection from the ionosphere into the magnetosphere. The storm conditions were significant enough to compress the magnetopause to a subsolar distance of ~6 $R_E$ and possibly closer. Under typical solar wind conditions, the expected geocentric distance of the magnetopause is ~10 $R_E$. Figure 1 shows that the MMS probes were on the inbound leg of their orbit. During this leg, the probes traversed the magnetosheath (MS), the magnetosheath boundary layer (MSBL), the magnetopause, the lower latitude boundary layer (LLBL), and the magnetosphere before the end of the mission-defined science region of interest (ROI)  (the science ROIs are defined approximately 3 months prior).

Comparing TOF spectra from each region is vital. Unfortunately, Burst Resolution TOF data, with the highest possible energy and temporal resolution (32 energies, 1-ns TOF resolution), is unavailable for all these intervals. However, an LLBL interval from 14:44:30 – 14:46:11 was acquired with the Burst Resolution. Due to its proximity to the magnetosphere and the fact that this boundary layer (as well as the MSBL) is produced by magnetic reconnection, the LLBL composition was of interest during this investigation. The burst resolution TOF-Flux plot and a cut at 3.2 keV from this interval are shown in Figure 3.

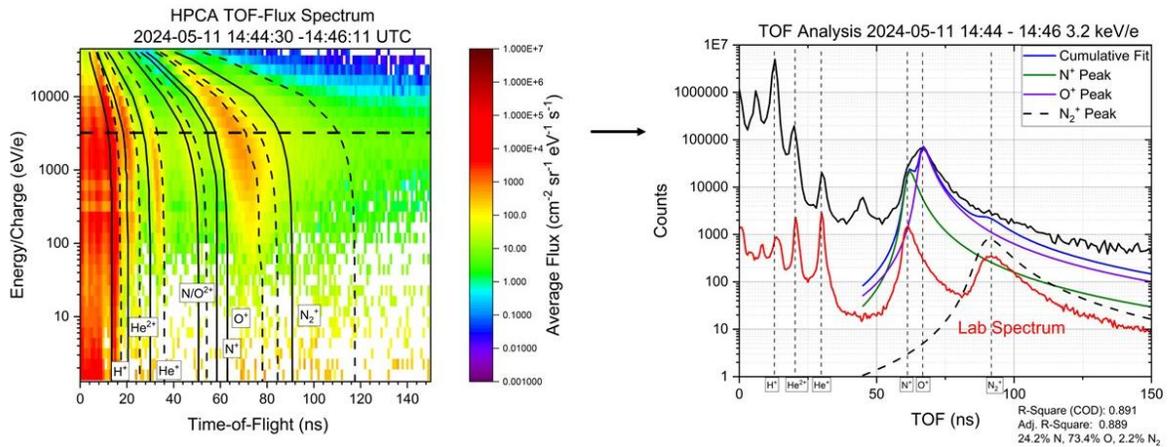

*Figure 3: HPCA TOF-Flux Spectrum 2024-05-11 14:44:30 - 14:46:11 UTC and the TOF spectrum cut at 3.2 keV/e. LEFT: TOF-Flux spectrum from LLBL. Includes ion contributions from atomic nitrogen, atomic oxygen, and molecular nitrogen, at lower levels than those in the magnetosphere. RIGHT: 3.2 keV/e cut showing the presence of oxygen and nitrogen—molecular nitrogen at a lower concentration than the magnetosphere.*

Measurements show that the LLBL is compositionally similar to the magnetosphere encounter, albeit with lower concentrations of the nitrogen ion species. Fitting the 3.2 keV/e cut spectrum yields relative concentrations of $N^+$, $O^+$, and $N_2^+$ of 24.2 , 73.4 %, and 2.2%. This result is unsurprising and even expected, given the proximity and the possibility of mixing between these regions. Overall, the heavy ion fluxes in the LLBL decreased relative to those in the magnetosphere by slightly more than an order of magnitude.

Estimating ion fluxes at a single energy does not provide a complete picture. Unlike the magnetosphere, the LLBL plasma is accelerated by the reconnection process. In the layer, as seen in Figure 1, the oxygen flux peaks at ~3 keV/e, where the 3.2 keV cut in Figure 3 occurs. If all ion populations in this layer propagate at the same velocity, independent of mass (Paschmann et al., 1989), then different ions will assume different energies. Oxygen ions at ~3 keV/e have an average velocity of ~200 km/s. Molecular nitrogen ions moving at this velocity have an energy of ~6 keV/e. A cut at this energy, taken from the same TOF-Flux spectrum shown in Figure 3, results in the TOF spectrum shown in Figure 4 (right).

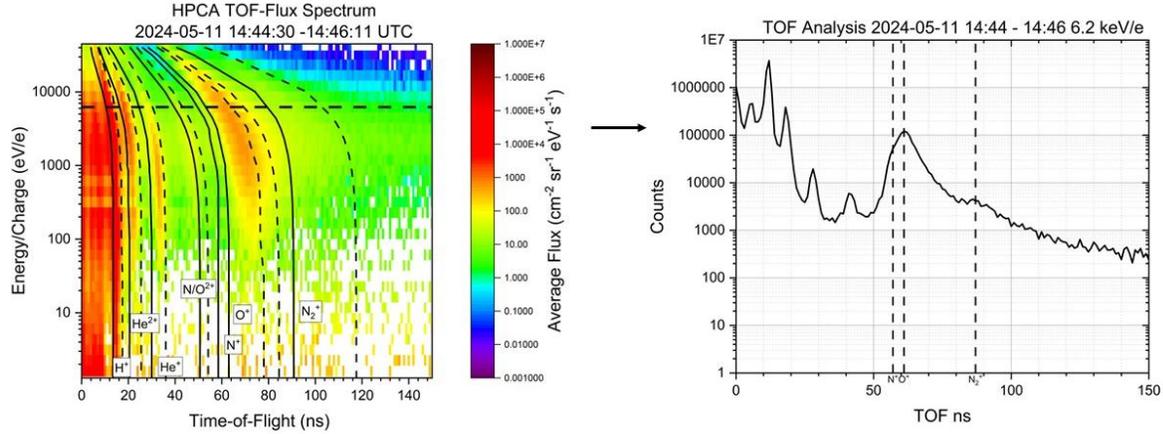

*Figure 4: LLBL TOF-Flux Spectrum and cut at 6.2 keV/e. LEFT: TOF-Flux spectrum from the LLBL encounter identical to the plot in Figure 3. RIGHT: TOF spectrum at 6.2 keV showing molecular nitrogen presence at comparatively higher energy.*

The plot displays, as expected, the presence of molecular nitrogen near the energy determined from the average ion velocity. Its relative height increases proportionally to the $N^+/O^+$ peak in the plot. Peak fits at this energy are not readily available since no laboratory calibration data exists. However, the importance of acquiring energy-dependent fits is not lost, and subsequent published work will investigate the relative concentrations of $N^+$, $O^+$, and $N_2^+$ over all energies in the reconnection boundary layers. What is evident from Figures 3 and 4 is that compositional analysis of the magnetosphere and boundary layers is possible with unprecedented temporal resolution. With the mass resolution, this temporal resolution allows a detailed study of the interplay between these regions, especially during geomagnetic storms, which are more frequent post-solar maximum.

## 7 Compositional Analysis and Computing the Reconnection Rate

One goal of the MMS-HPCA investigation was to determine the effect of massive ions on the process of magnetic reconnection (Burch et al., 2015). Until now, the only ions considered were those immediately identified in the HPCA ion products. The only heavy ion identified in these products is atomic oxygen. However, the TOF spectra in Figure 2 and Figure 3 show that $O^+$ is not the only heavy ion species present, so its contributions are only a partial measure of the overall effect.

Cassak & Shay (2007) and Birn et al. (2008) provide the equation (1) used to determine the reconnection rate:

$$R = \frac{(\rho_s B_M)^{1/2}}{(\rho_M B_s + \rho_s B_M)^{1/2}} \tag{1}$$

Using the notation from Borovsky et al. (2013), R is the fractional reduction of the local dayside reconnection rate due to the nonzero magnetospheric ion populations. The mass density, $\rho$, includes the masses of all ions present. B is the strength of the magnetic field's reconnecting component in the magnetosphere (subscript – s) and the magnetosheath (subscript – M). Fuselier et al. (2017; 2019) used Equation 1 to estimate the change in the reconnection rate in the presence of $O^+$ for nominal conditions and for moderate storms, assuming all heavy ions were $O^+$.

HPCA, as stated previously, converts TOF data into ion velocity distribution functions in four ranges corresponding to $H^+$, $He^{2+}$, $He^+$, and $O^+$. Then, the moment values of the respective VDFs (ion number density, ion velocity, ion scalar and vector temperatures, and the ion pressure tensor) are determined. The ion number densities are the pertinent data products for calculating the reconnection rate using Equation 1. The average ion number densities for the magnetopause and magnetosphere are shown in the table below. The last column in the table uses the relative nitrogen, oxygen, and molecular nitrogen peak measurements to estimate the respective number densities based on HPCA's oxygen density measurement.

*Table 1: HPCA Ion Densities Used for Assessing Reconnection Rate Reduction via Mass Loading*

| Ion Species | Number Density- Magnetosheath (cm$^{-3}$) | Number Density – Magnetosphere (cm$^{-3}$) - oxygen only | Number Density – Magnetosphere (cm$^{-3}$) -O+, N+, N2+ |
|---|---|---|---|
| $H^+$ | 30.72 ± 1.86 | 2.35 ± 0.64 | 2.35 ± 0.64 |
| $He^{2+}$ | 3.63 ± 0.27 | 0.23 ± 0.07 | 0.23 ± 0.07 |
| $He^+$ | 0 | 0.10 ± 0.01 | 0.10 ± 0.01 |
| $O^+$ | 0 | 6.45 ± 0.50 | 3.67 ± 0.27 |
| $N^+$ | 0 | 0 | 2.77 ± 0.22 |

| $N_2^+$ | 0 | 0 | 0.53 ± 0.04 |

For these intervals, it's assumed that the reconnecting component of the magnetic field is $B_z$. At the magnetopause, this component is 119.5 ± 6.08 nT, while in the magnetosphere, it is 128.5 ± 3.42 nT. Using only the ion-specific density values shown in the table and the magnetic field values in the magnetosheath and magnetosphere, Equation 1 yields a fractional reduction in the reconnection rate of 0.560. That is, the rate is reduced by 44% because of the presence of $O^+$. This reconnection rate reduction doesn't include contributions from any heavy ion besides oxygen.

Using data from the TOF spectrum in the magnetosphere, it's evident that other heavy ions are also present. With the relative abundance values determined from the fits and assuming that these proportions remain constant throughout the interval, the number densities for $N^+$, $O^+$, and $N_2^+$ become 2.77 ± 0.22 cm-3, 3.67 ± 0.27 cm-3, and 0.53 ± 0.04 cm-3 respectively, where the measurement uncertainty is distributed with the measured values. When these densities are included in Equation 1, the resulting reconnection rate is 0.544, a 46% reduction. There is little difference between the two estimates of the decrease in the reconnection rate because while $N^+$ is less massive than $O^+$, the presence of $N_2^+$ offsets the mass loss due to the decreased oxygen. Hence, the mass densities in the magnetosphere in Equation 1 are similar. Such a slight change in the reduction in the reconnection rate is essentially impossible to measure from the observations.

## 8 Conclusions

The Gannon Storm yielded enormous opportunities for scientific study in multiple regions of the Earth's magnetosphere. The external pressure caused by the storm compressed the magnetosphere to a geocentric radius of ~ 6 $R_E$ and caused the injection of ionospheric species into the nightside magnetosphere. Ultimately, these ions were likely accelerated in the tail and eventually convected to the dayside, given the peak energy at which they were observed. The Magnetospheric Multiscale Mission Hot Plasma Composition Analyzers traversed the dayside magnetosphere during the early recovery storm phase and provided the first measurement of these species, namely atomic and molecular nitrogen in the dayside magnetopause near the subsolar point. In the supplemental material, a TOF spectrum from the MSBL interval shows that N2+ escapes along open field lines into the magnetosheath.

The presence of more massive ion species in the magnetosphere leads to mass loading, which adds to the plasma inertia. Such mass loading is expected to reduce the reconnection rate, as predicted from Equation 1. It's important to point out that while calculating a true reconnection rate is difficult, the results of this short study point out that reconnection persists regardless of the storm conditions, albeit at a slower rate than it would in the absence of storm-caused perturbation. The observations and measurements in the LLBL and MSBL also support this conclusion.

Assuming that the compositional proportions in the magnetosphere at 3.2 keV/e remained constant throughout the interval is significant but not necessarily unfounded, given the ion distributions of these ions over the instrument's energy range. In-depth analysis of the TOF spectra requires modifying the fits relative to energy. While plausible, such analysis is not in the scope of this work.

One more item of note is the presence of what appears to be ions with an M/Q = 7 or 8, identifying them as either doubly charged oxygen or nitrogen. The difficulty in identifying these ions lies mainly in the peak fitting. Producing doubly ionized species in a calibration facility is notoriously difficult, and therefore, no doubly ionized species were used in the lab calibration. The same difficulty does not manifest in space, and $O^{2+}$ observations occur regularly during MMS orbits.

Figure 2 and Figure 3 show that ions that track along the expected TOF range of M/Q = 8 but trend with shorter flight times. Shorter times would identify these ions as doubly-charged nitrogen atoms. Given that nitrogen's second ionization energy (2856 kJ/mol) is lower than oxygen's (3388.3 kJ/mol) and the enhanced nitrogen supply available during this period, these ions may be M/Q = 7, which also makes this the first magnetospheric observation of these ions on the dayside as well. These observations, the increased incidence of storms, and HPCA's measurement capabilities allow short- and long-term investigations of storm effects on this portion of the Earth's magnetosphere. These measurements will shed light on the paths these heavy ions take from the ionosphere to the dayside magnetosphere and contribute significantly to space weather studies.

10   Open Research

MMS data used in this article are publicly available at the MMS Science Data Center (https://lasp.colorado.edu/mms/sdc/public/).

11   Supplementary Material – The Survey Time of Flight Data Product

When not in the mission-defined science regions of interest (ROIs), HPCA still acquires time-of-flight data, albeit with reduced TOF and energy resolution. This reduced, or survey mode, TOF product is first used to produce the ion-specific fluxes, velocity distributions, and their associated moments (ion number density, ion velocity, etc.). Once processed, the TOF product is reduced by a factor of 4 in energy and a TOF time resolution of two nanoseconds, i.e., 16 E/q channels by 256 TOF. However, the counts in adjacent TOF channels are coadded and then redistributed evenly. Thus, two adjacent channels, 0 & 1, 2 & 3, etc., have the same counts allocated to them. This reallocation causes the TOF spectrum to have a "blocky" appearance. In addition, the TOF-Flux spectrum appears to have a higher flux level than its burst resolution counterpart. The following Figure shows survey plots from an encounter with the Magnetosheath Boundary Layer (MSBL)

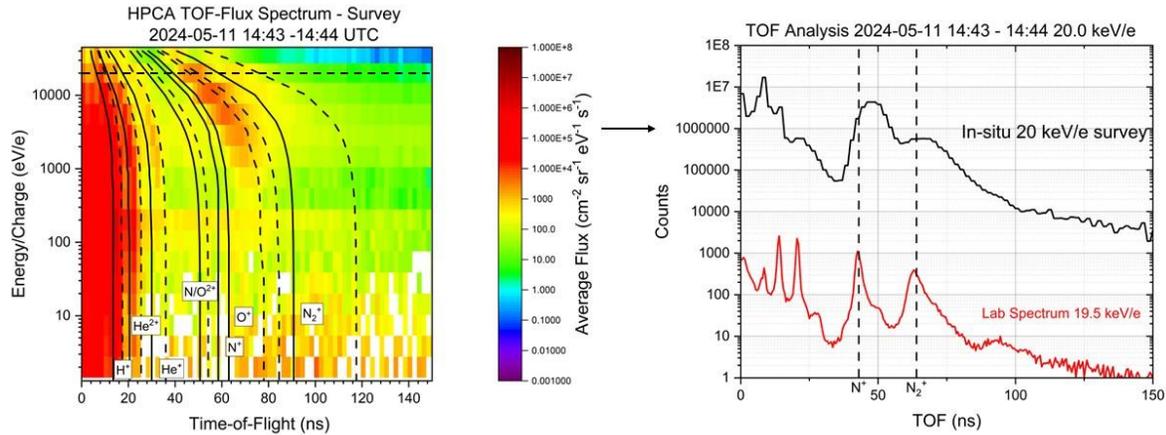

*Figure 5:Survey resolution TOF product from the MSBL encounter. On the left, the TOF-Flux plot displays a higher background and maximum flux for the period. However, these effects are easily attributable to the co-addition of energy and TOF channels used to reduce the size of the stored product. At the right is a TOF cut at 20 keV/e for in-situ data compared to a calibration spectrum at 19.5 keV/e. The rounded appearance of the in-situ spectrum is due to the co-addition of TOFs from four adjacent energy steps.*

As shown in the plot, flux calculations and qualitative measurements of the ion species present are possible using the survey TOF product. The Figure shows that fitting survey resolution spectra presents a specific challenge: peak broadening. In survey mode, counts from four adjacent energies are added. At lower energies, where the TOF has a relaxed dependence on energy (below 1 keV/e in Figure 5), fitting a peak with contributions from several energies is straightforward. However, at higher energies, for instance, the 20 keV/e cut in the Figure, the TOF changes sharply with E/q. Thus, the peak from the four energy steps is broad and appears almost Gaussian. The TOF spectrum acquired during calibration (the red trace in the Figure at 19.5 keV/e) shows how well HPCA resolves the atomic and molecular nitrogen ions.

The 20 keV/e cut is interesting because it shows molecular nitrogen escaping from the MSBL. During 14:43 – 14:44, the z-component of the magnetic field is directed steadily southward, while the protons move opposite that direction, northward. Again, assuming ions are accelerated to the same velocity, molecular nitrogen is expected to appear at an energy that's about twenty-eight times greater, near 20 keV/e. The cut at this energy shows a clear $N_2^+$ signature, which indicates molecular nitrogen escaping from the MSBL, another first.